\documentclass{jfm} %arXiv does not like line numbering

\usepackage{graphicx}
\usepackage{newtxtext}
\usepackage{newtxmath}
\usepackage{natbib}
\usepackage{hyperref}
\usepackage{subcaption}
\usepackage{wasysym}
\hypersetup{
    colorlinks = true,
    urlcolor   = blue,
    citecolor  = black,
}

\newcommand{\RomanNumeralCaps}[1]
\linenumbers

\usepackage{soul}
\usepackage{adjustbox}
\usepackage[dvipsnames]{xcolor}
\usepackage{colortbl}
\usepackage[dvipsnames]{xcolor}
\usepackage{mathtools,xparse}
\usepackage{lipsum}
\usepackage{bm}

\usepackage[ruled,vlined]{algorithm2e}
\SetAlCapHSkip{0pt}

\DeclarePairedDelimiter{\abs}{\lvert}{\rvert}
\DeclarePairedDelimiter{\norm}{\lVert}{\rVert}
\DeclarePairedDelimiter{\binner}{\big\langle}{\big\rangle}

\DeclarePairedDelimiterX{\dotp}[2]{\big\langle}{\big\rangle}{#1, #2}

\newcommand{\eps}{\varepsilon}
\newcommand{\datacov}{\mathcal{C}_{\bm{u}^\star}}
\newcommand{\priorcov}{\mathcal{C}_{\bar{\bm{x}}}}
\newcommand{\postcovmap}{\mathcal{C}_{\bm{x}^\circ}}
\newcommand{\postcovxk}{\mathcal{C}_{\bm{x}_k}}

\newcommand{\figwidth}{0.45}
\newcommand{\figwidthh}{0.8}
\title{Learning rheological parameters of non-Newtonian fluids from velocimetry data}

\author{Alexandros Kontogiannis\aff{1}
  \corresp{\email{ak2239@cam.ac.uk}},
  Richard Hodgkinson\aff{2},
  Steven Reynolds\aff{3}
 \and Emily L. Manchester\aff{4}}

\affiliation{\aff{1}Engineering Department, University of Cambridge, Trumpington Street, Cambridge, CB2 1PZ, UK
\aff{2}Materials Science and Engineering Department, University of Sheffield, Sheffield, S1 3JD, UK
\aff{3}School of Medicine and Population Health, Faculty of Health, University of Sheffield, Sheffield, S10 2RX
\aff{4}Mechanical \& Aerospace Engineering Department, University of Manchester, Manchester, M13 9PL, UK}

\begin{document}
\maketitle

\begin{abstract}
We solve a Bayesian inverse Navier--Stokes (N--S) problem that assimilates velocimetry data by jointly reconstructing a flow field and learning its unknown \mbox{N--S} parameters. We devise an algorithm that learns the most likely parameters of a Carreau shear-thinning viscosity model, and estimates their uncertainties, from velocimetry data of a shear-thinning fluid. We conduct a flow-MRI experiment to obtain velocimetry data of an axisymmetric laminar jet in an idealised medical device (FDA nozzle) for a blood analogue fluid. The algorithm successfully reconstructs the flow field and learns the most likely Carreau parameters. Predictions from the learned model agree well with rheometry measurements. The algorithm accepts any differentiable algebraic viscosity model, and can be extended to more complicated \mbox{non-Newtonian fluids} (e.g. Oldroyd-B fluid if a viscoelastic model is incorporated).
\end{abstract}


% {\bf MSC Codes }  {\it(Optional)} Please enter your MSC Codes here

\section{Introduction}
% flow-MRI and its limitations (rheology)
Magnetic resonance velocimetry (flow-MRI) is an experimental technique that measures fluid velocities in time and three-dimensional space. Flow-MRI is most commonly known for \textit{in vivo} clinical settings but is gaining popularity within the wider scientific community for \textit{in vitro} applications \citep{Elkins2007}. Whilst flow-MRI can provide reliable velocity measurements, it does not directly provide information about fluid properties such as rheology or pressure. These currently require additional experiments to
measure a fluid’s shear stress-strain curve. Acquiring this non-invasively is challenging as it requires knowledge about both the stress and strain, and some control of either. Common experimental techniques to measure  fluid viscosity include rotational and capillary rheometry, which involve passing a fluid sample through a precice geometry and measuring shear rate, torque, or pressure drop. {\color{black}Other techniques are available such as industrial `in-line' and `on-line' rheometry, or ultrasound velocity profiling (UVP). However, these methods are either highly invasive \citep{KonigsbergNicholsonHalleyKealyBhattacharjee+2013} or may require pressure drop measurements \citep{Krishna2022}. Due to the additional costs and complexities of rheometry experiments, it is not always feasible to acquire rheological data. A non-intrusive in-line UVP rheometry method that does not require pressure drop measurements is presented in \cite{Tasaka2021}. More recent techniques include ultrasonic \citep{Yoshida2022,Ohie2022} and optical \citep{Noto2023} spinning rheometry.}

For computational fluid dynamic (CFD) simulations of non-Newtonian fluids, rheological behaviour is expressed through viscosity models with adjustable parameters. Model parameters are typically taken from the literature. 
%Frame this as a potential application of the work
In biomedical engineering, flow-MRI data can inform patient-specific cardiovascular models. Without patient-specific blood rheology, CFD models lack accuracy. \cite{Ranftl2023} performed uncertainty quantification to investigate the impact of non-Newtonian and Newtonian CFD models on haemodynamic outputs. They found that patient rheological properties are necessary for accurate wall shear stress predictions, particularly for diseases where blood properties differ from those in healthy populations, and in small arteries where non-Newtonian effects dominate.

% Lit review
Bayesian inference is a data-driven technique that can estimate unknown physical or model parameters and their uncertainties from experimental data combined with some prior knowledge. \cite{Worthen2014} inferred two unknown parameters of a constitutive equation for viscosity in mantle flow with this approach. The forward problem was governed by a nonlinear Stokes problem and experimental data was of surface velocity measurements. Their method recovered constant and spatially-varying parameters reasonably well. Our conceptual approach is similar although the application and technical details differ.

%Summary
In this study we infer the rheological parameters of a shear-thinning blood analogue from flow-MRI-measured \emph{velocity fields} alone. We select the Carreau model to represent the non-Newtonian fluid behaviour \citep{Sequeira2007} because it is differentiable and bounded. Experiments are performed on the Food and Drug Administration's (FDA) benchmark nozzle, and data is assimilated using a Bayesian inverse Navier--Stokes problem that jointly reconstructs the flow field and learns the Carreau parameters. {\color{black}Our inversion algorithm differs from the aforementioned rheometry methods by inferring rheological properties of non-Newtonian fluids from \emph{general velocity fields} alone (as long as information on the viscous stress tensor can be retrieved from the field). It relies solely on velocity field measurements, which can be acquired using any velocimetry technique (e.g. flow-MRI, Doppler, or particle image velocimetry)}. This is the first time flow-MRI has been used for a non-invasive measurement of rheological parameters.

\section{Bayesian inversion of the Navier--Stokes problem}
\label{sec:learning_rheo}

We learn the rheology of a non-Newtonian fluid from velocimetry data by solving a \emph{Bayesian inverse \mbox{N--S} problem}. We first assume that there is a N--S problem with a Carreau fluid model that can explain the velocimetry data, $\bm{u}^\star$. Therefore N--S parameters, $\bm{x}^\circ$, exist such that
\begin{gather}
\bm{u}^\star -\mathcal{Z}\bm{x}^\circ = \bm{\eps} \sim \mathcal{N}(\bm{0},\datacov)\quad,
\end{gather}
where $\mathcal{Z}$ is the nonlinear operator that maps N--S parameters to N--S solutions projected into the data space, and $\bm{\eps}$ is Gaussian noise with zero mean and covariance operator $\datacov$. We do not know $\bm{x}^\circ$, but we assume that its prior probability distribution is $\mathcal{N}(\bar{\bm{x}},\priorcov)$, where $\bar{\bm{x}}$ is the prior mean, and $\priorcov$ is the prior covariance operator. Using Bayes' theorem we then find that the posterior probability density function (p.d.f.) of $\bm{x}$, given the data $\bm{u}^\star$, is given by
\begin{equation}
\pi\big(\bm{x}\big|\bm{u}^\star\big) \propto \pi\big(\bm{u}^\star\big|\bm{x}\big)~\pi(\bm{x}) = \exp\Big(-\frac{1}{2}\norm{\bm{u}^\star-\mathcal{Z}\bm{x}}^2_{\datacov} 
- \frac{1}{2}\norm{\bm{x}-\bar{\bm{x}}}^2_{\priorcov}\Big)
\label{eq:posterior_pdf}\quad,
\end{equation}
where $\pi\big(\bm{u}^\star\big|\bm{x}\big)$ is the data likelihood, $\pi\big(\bm{x}\big)$ is the prior p.d.f. of $\bm{x}$, and $\norm{\cdot,\cdot}^2_{\mathcal{C}} \coloneqq \binner{\cdot,\mathcal{C}^{-1}\cdot}$ is the covariance-weighted $L^2$-norm. The most likely parameters, $\bm{x}^\circ$, maximise the posterior p.d.f. (maximum \emph{a posteriori} probability, or MAP estimator), and are given implicitly as the solution of the nonlinear optimisation problem
\begin{gather}
\quad \bm{x}^\circ \equiv \underset{\bm{x}}{\mathrm{argmin}}\mathscr{J} \quad,\quad \text{where}\quad \mathscr{J} \coloneqq \frac{1}{2}\norm{\bm{u}^\star-\mathcal{Z}\bm{x}}^2_{\datacov} 
+ \frac{1}{2}\norm{\bm{x}-\bar{\bm{x}}}^2_{\priorcov}\quad.
\label{eq:map_opt_problem}
\end{gather}

Using a first order Taylor expansion of $\mathcal{Z}$ around $\bm{x}_k$, given by
\begin{gather}
\mathcal{Z}\bm{x} \simeq \mathcal{Z}\bm{x}_k + \mathcal{G}_k~\big(\bm{x}-\bm{x}_k)\quad,
\label{eq:z_taylor_exp}
\end{gather}
the optimality conditions of problem \eqref{eq:map_opt_problem} lead to the following iteration \citep[Chapter~6.22.6]{Tarantola2005}
\begin{gather}
\bm{x}_{k+1} \mapsfrom \bm{x}_k - \tau_k~\postcovxk~ \big(D_{\bm{x}}\mathscr{J}\big)_k\quad,
\label{eq:map_update_rule}
\end{gather}
where 
\begin{gather}
\big(D_{\bm{x}}\mathscr{J}\big)_k \coloneqq -\mathcal{G}_k^*~\datacov^{-1}\big(\bm{u}^\star-\mathcal{Z}\bm{x}_k\big) + \priorcov^{-1}\big(\bm{x}_k-\bar{\bm{x}}\big)\quad,
\label{eq:map_grad}
\end{gather}
$\mathbb{R}\ni\tau_k > 0$ is the step size at iteration $k$, which is determined by a line-search algorithm, $\mathcal{G}^{*}_k$ is the adjoint of $\mathcal{G}_k$, and $\postcovxk$ is the posterior covariance operator at iteration $k$, which is given by
\begin{gather}
\postcovxk \coloneqq \big(\mathcal{G}_k^*~\datacov^{-1}~\mathcal{G}_k+\priorcov^{-1}\big)^{-1}\quad.
\label{eq:map_cov}
\end{gather}
The posterior covariance operator around the MAP estimate, $\mathcal{C}_{\bm{x}^\circ}$, can then be used to approximate the posterior p.d.f. such that
\begin{equation}
\pi\big(\bm{x}\big|\bm{u}^\star\big)\simeq \exp\Big(-\frac{1}{2}\norm{\bm{x}-\bm{x}^\circ}^2_{\postcovmap} - \textrm{const.}\Big)\quad,
\label{eq:laplace_approx}
\end{equation}
which is known as the \emph{Laplace approximation} \citep[Chapter~27]{MacKay2003}. For linear models, the approximation is exact when both $\pi\big(\bm{u}^\star\big|\bm{x}\big)$ and $\pi(\bm{x})$ are normal. For nonlinear models, the accuracy of the approximation depends on the behaviour of the operator $\mathcal{Z}$ around the critical point $\bm{x}^\circ$ \citep[Section~2.2]{Kontogiannis2024}. 

\subsection{N--S problem and the operators $\mathcal{Z}$, $\mathcal{G}$}
In order to solve the inverse problem \eqref{eq:map_opt_problem} using formulas \eqref{eq:map_update_rule}-\eqref{eq:map_cov}, we need to define $\mathcal{Z}$ and $\mathcal{G}$. We start from the N--S boundary value problem in $\Omega \subset \mathbb{R}^3$
\begin{gather}
\bm{u}\bm{\cdot}\nabla\bm{u}-\nabla\bm{\cdot}\big(2{\color{black}\nu_e}\nabla^s\bm{u}\big)  + \nabla p = \bm{0} \quad\text{and}\quad \nabla \bm{\cdot} \bm{u} = 0 \quad \textrm{in}\quad {\color{black}\Omega}\quad,\nonumber\\
\bm{u} = \bm{0} \quad \textrm{on}\quad \Gamma \quad,\quad \bm{u} = {\color{black}\bm{g}_i} \quad \textrm{on}\quad\Gamma_i\quad,\quad -2{\color{black}\nu_e}\nabla^s\bm{u}\bm{\cdot}\bm{\nu}+p\bm{\nu} = {\color{black}\bm{g}_o} \quad \textrm{on} \quad \Gamma_o\quad,
\label{eq:navierstokes_bvp}
\end{gather}
where $\bm{u}$ is the velocity, $p\mapsfrom p/\rho$ is the reduced pressure, $\rho$ is the density, $\nu_e$ is the \emph{effective} kinematic viscosity, $(\nabla^s \bm{u})_{ij} \coloneqq \frac{1}{2}(\partial_j u_i + \partial_i u_j)$ is the strain-rate tensor, $\bm{g}_i$ is the Dirichlet boundary condition (b.c.) at the inlet $\Gamma_i$, $\bm{g}_o$ is the natural b.c. at the outlet $\Gamma_o$, and $\bm{\nu}$ is the unit normal vector on the boundary $\partial\Omega = \Gamma\cup\Gamma_i\cup\Gamma_o$, where $\Gamma$ is the no-slip boundary (wall). The construction of the operators $\mathcal{Z}, \mathcal{G}$ of the generalised inverse N--S problem, whose unknown parameters are the shape of the domain $\Omega$, the boundary conditions $\bm{g}_i, \bm{g}_o$, and the viscosity field $\nu_e$, is treated in \cite{Kontogiannis2021,Kontogiannis2022b,Kontogiannis2024}. Here, we fix the geometry, $\Omega$, and the outlet b.c., $\bm{g}_o$, and infer only the inlet b.c., $\bm{g}_i$, and the effective viscosity field, $\nu_e$. We further introduce the Carreau model for the effective viscosity field, which is given by
\begin{align}
\mu_e(\dot{\gamma};{\color{black}\bm{p}_\mu}) &\coloneqq {\color{black}\mu_\infty} + {\color{black}\delta\mu}\big(1 +({\color{black}\lambda}\dot{\gamma})^2\big)^{({{\color{black}n}-1})/{2}}\quad,
\label{eq:carreau_model}
\end{align}
where $\mu_e \coloneqq \nu_e\rho$, $\dot{\gamma}(\bm{u}) \coloneqq \sqrt{2\nabla^s\bm{u}\bm{:}\nabla^s\bm{u}}$ is the magnitude of the strain-rate tensor, and \mbox{$\bm{p}_\mu\coloneqq\big(\mu_\infty,\delta\mu,\lambda,n\big)$} are the Carreau fluid parameters. In order to infer the most likely viscosity field, $\mu^\circ_e$, we therefore need to infer the most likely Carreau fluid parameters, $\bm{p}^\circ_\mu$. 

After linearising problem \eqref{eq:navierstokes_bvp} around $\bm{u}_k$, we obtain $\bm{u}(\bm{x}) \simeq \bm{u}_k + \mathcal{A}_k \big(\bm{x}-\bm{x}_k)$, where $\mathcal{A}_k \equiv \big((D^{\mathscr{M}}_{\bm{u}})^{-1}D^{\mathscr{M}}_{\bm{x}}\big)_k$, with $\mathcal{A}_k$ being a linear, \emph{invertible} operator, which encapsulates the inverse Jacobian of the N--S problem, $(D^{\mathscr{M}}_{\bm{u}})^{-1}$, and the generalised gradient of the velocity field with respect to the parameters $\bm{x}$, $D^{\mathscr{M}}_{\bm{x}}$. Observing that $\mathcal{Z}, \mathcal{G}$ map from the N--S parameter space to the (velocimetry) data space, we define $\mathcal{Z} \coloneqq \mathcal{S}\mathcal{Q}$, and $\mathcal{G}_k \coloneqq \mathcal{S}\mathcal{A}_k $, where $\mathcal{S}:\bm{M}\to \bm{D}$ is a projection from the model space, $\bm{M}$, to the data space, $\bm{D}$, and $\mathcal{Q}$ is the operator that maps $\bm{x}$ to $\bm{u}$, \emph{i.e.} that solves the N--S problem. (The operators $\mathcal{S},\mathcal{Q},\mathcal{A}$ are derived in \cite{Kontogiannis2024} from the weak form of the N--S problem \eqref{eq:navierstokes_bvp}, $\mathscr{M}$.)

Based on the above definitions, and due to \eqref{eq:map_grad}, we observe that the \emph{model contribution} to the objective's steepest descent direction, for the Carreau parameters, $\bm{p}_\mu$, is
\begin{gather}
\big(\delta\bm{p}_{\mu}\big)_k \coloneqq \mathcal{G}_k^*~\datacov^{-1}\big(\bm{u}^\star-\mathcal{Z}\bm{x}_k\big) = \underbrace{\big(D^{\mathscr{M}}_{\bm{p}_{\mu}}\big)^*_k~ \big((D^{\mathscr{M}}_{\bm{u}})^{*}\big)^{-1}_k}_{D^{\bm{p}_{\mu}}_{\bm{u}}}~\underbrace{\mathcal{S}^*~\datacov^{-1}\big(\bm{u}^\star-\mathcal{S}\bm{u}_k\big)}_{\text{data-model discrepancy } \delta\bm{u} \in \bm{M}}\quad.
\label{eq:carreau_steepest_descent_compact}
\end{gather}
Even though $D^\mathscr{M}_{\bm{u}}$ is invertible, for large-scale problems (such as those in fluid dynamics) its inverse, $(D^\mathscr{M}_{\bm{u}})^{-1}$, cannot be stored in computer memory because its discrete form produces a dense matrix. The discrete form of $D^\mathscr{M}_{\bm{u}}$, however, produces a sparse matrix. Consequently, instead of using the explicit formula \eqref{eq:carreau_steepest_descent_compact}, the steepest descent directions are given by
\begin{gather}
\big(\delta\bm{p}_{\mu}\big)_k \coloneqq \big(D^{\mathscr{M}}_{\bm{p}_{\mu}}\big)^*_k~\bm{v}_k = 2\int_\Omega\big(D_{\bm{p}_\mu}\mu_e\big)_k\big(\nabla^s\bm{u}_k\bm{:}\nabla^s\bm{v}_k\big)\quad,
\label{eq:adjoint_based_grad}
\end{gather}
where $\big(D_{\bm{p}_\mu}\mu_e\big)_k \equiv D_{\bm{p}_\mu}\mu_e\big(\dot{\gamma}(\bm{u}_k);(\bm{p}_{\mu})_k\big)$ consists of the derivatives of the Carreau model with respect to its parameters, and $\bm{v}_k$ is the adjoint velocity field, which is obtained by solving the following linear operator equation  
\begin{gather}
A\bm{v}_k = b \quad,\quad \text{where}\quad A \equiv \big(D^{\mathscr{M}}_{\bm{u}}\big)_k^* \quad,\quad \text{and}\quad b \equiv \mathcal{S}^*\datacov^{-1}\big(\bm{u}^\star-\mathcal{S}\bm{u}_k\big)\quad.
\label{eq:adjoint_ns_prob}
\end{gather}
The steepest descent directions for the inlet b.c., $\bm{g}_i$, are derived in \cite{Kontogiannis2024}. 

Instead of explicitly computing $\postcovxk$ at every iteration using formula \eqref{eq:map_cov}, we approximate $\postcovxk$ using the damped BFGS quasi-Newton method \citep{Goldfarb2020}, which ensures that $\postcovxk$ remains positive definite, and its approximation remains numerically stable.
\begin{algorithm}[h]
\textbf{Input:} velocimetry data, $\bm{u}^\star$, data cov., $\datacov$, prior mean, $\bar{\bm{x}}$, and prior cov. $\priorcov$\\
\textbf{Initialisation:} set $k \mapsfrom 0$, $\bm{x}_0 \mapsfrom \bar{\bm{x}}$, and compute initial velocity field $\bm{u}_0 \mapsfrom \mathcal{Q}~\bm{x}_0$\\
\While{$\mathtt{termination\_criterion\_is\_not\_met}$}{
$\makebox[0pt][l]{$\bm{v}_k$}\phantom{\hspace{1.75cm}} \mapsfrom$~\makebox[0pt][l]{solve adjoint N--S problem around $\bm{u}_k$}\phantom{\hspace{6cm}}\quad(eq.~\eqref{eq:adjoint_ns_prob})\\
$\makebox[0pt][l]{$\big(D_{\bm{x}}\mathscr{J}\big)_k$}\phantom{\hspace{1.75cm}} \mapsfrom$~\makebox[0pt][l]{compute steepest descent directions}\phantom{\hspace{6cm}}\quad(eq.~\eqref{eq:map_update_rule},~\eqref{eq:adjoint_based_grad})\\
$\makebox[0pt][l]{$\postcovxk, \tau_k$}\phantom{\hspace{1.75cm}} \mapsfrom$~\makebox[0pt][l]{update post. cov. approx. and find step size}\phantom{\hspace{6cm}}\quad(damped BFGS)\\
$\makebox[0pt][l]{$\bm{x}_{k+1}$}\phantom{\hspace{1.75cm}} \mapsfrom$~\makebox[0pt][l]{$\bm{x}_k - \tau_k~\postcovxk D_{\bm{x}}\mathscr{J}$, \emph{i.e.} update N--S parameters}\phantom{\hspace{6cm}}\\
$\makebox[0pt][l]{$\bm{u}_{k+1}$}\phantom{\hspace{1.75cm}} \mapsfrom$~\makebox[0pt][l]{$\mathcal{Q}~\bm{x}_{k+1}$, \emph{i.e.} update N--S solution, and set $k\mapsfrom k+1$}\phantom{\hspace{10cm}}\\
% 6. $\makebox[0pt][l]{$\mathscr{J}_{k+1}$}\phantom{\hspace{1.75cm}} \mapsfrom$ \makebox[0pt][l]{compute objective function and set $k\mapsfrom k+1$}\phantom{\hspace{10cm}} \quad(eq. \eqref{eq:posterior_pdf})
}%end while
\textbf{Output:} \emph{MAP estimates}: $\bm{u}^\circ \mapsfrom \bm{u}_k$, $p^\circ \mapsfrom p_k$, $\bm{x}^\circ \mapsfrom \bm{x}_k$ and post. covariance $\mathcal{C}_{\bm{x}^\circ}$ 
\caption{Learning rheological parameters from velocimetry data}\label{algo_learning_rheo}
\end{algorithm}
% }
% \end{adjustwidth}}

{\color{black}
\subsubsection{Note on effective viscosity model selection}
In this study, although we fix the effective viscosity model to the Carreau fluid model, which is given by equation \eqref{eq:carreau_model}, the present Bayesian inversion framework is already set up for model selection \cite[Section~4.3]{Yoko_Juniper_2024b}. The velocimetry data, $\bm{u}^\star$, can be assimilated into the N--S boundary value problem with as many different viscosity models as the user likes. The model parameter uncertainties are then estimated, and the marginal likelihood of each model is calculated. The marginal likelihood is the likelihood of each model, given the experimental data, which is also known as the evidence for each model. The models are then ranked by their evidence, and the most accurate model is chosen. Bayesian rheological model ranking using \emph{rheometry} data has been addressed in \cite{10.1122/1.4915299}. Note that, here we infer the rheology of the fluid from \emph{velocimetry} data (instead of rheometry data), which is a more difficult problem. Bayesian rheological model ranking from \emph{velocimetry} data is a natural extension of the present Bayesian inversion framework\footnote{\color{black}Another extension that follows naturally is optimal experiment design \citep[Section~4.1]{Yoko_Juniper_2024}, i.e. strategically planning experiments to maximise information gain in parameter estimation.}, and provides scope for future work.
}

\section{Flow-MRI experiment of a non-Newtonian laminar jet}
\begin{figure}
\centering
\includegraphics[width=0.8\textwidth]{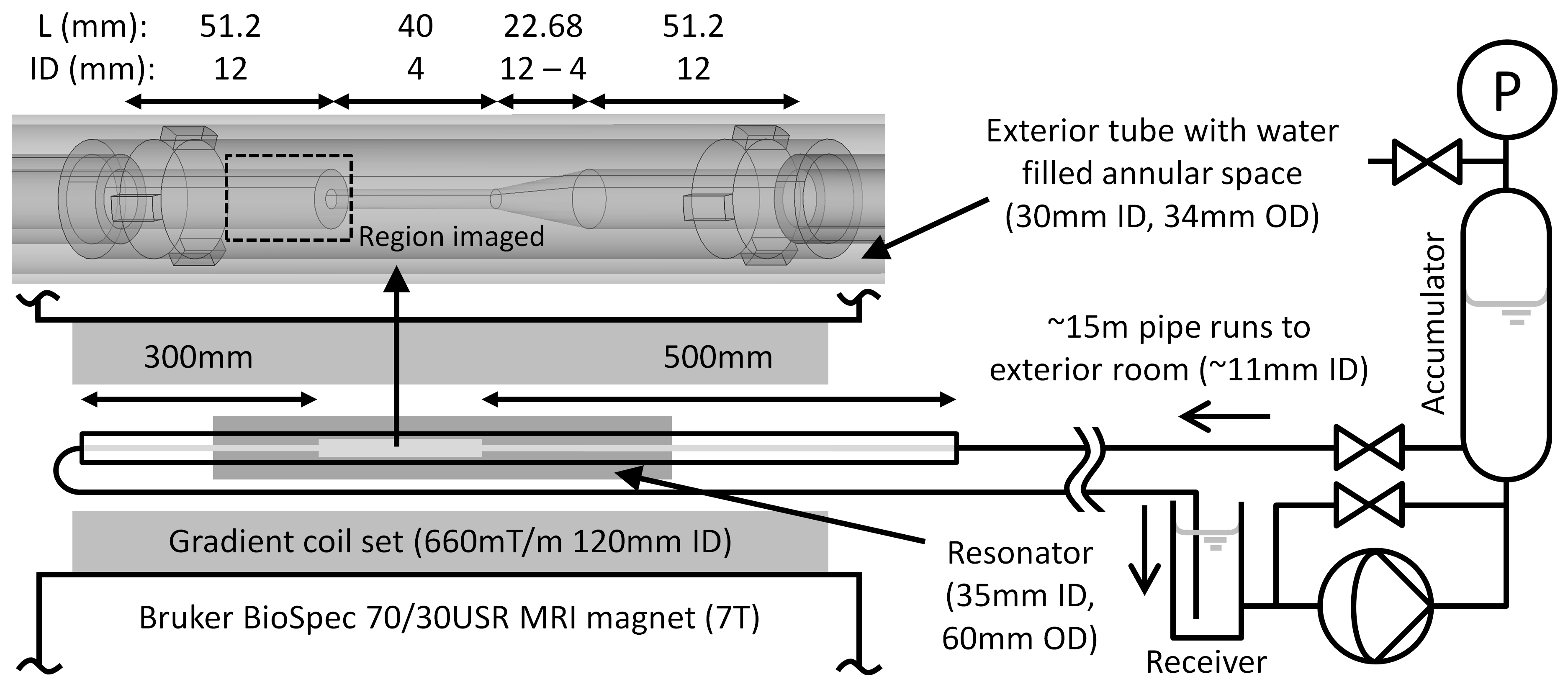}
\caption{Overall flow system and setup around the MRI scanner with detail of the FDA flow nozzle geometry implemented. ID: Inner diameter, OD: outer diameter.}
\label{fig:mri_setup}
\end{figure}

% 1. experiment setup (hydraulics)
The test section is part of the FDA nozzle \citep{fda_nozzle,Stewart2012}, which is an axisymmetric pipe that converges to a narrow throat section, followed by a sudden expansion, where a non-Newtonian laminar jet forms (see figure \ref{fig:mri_setup}). The geometry was 3D-resin-printed to a nominal tolerance of $\pm$0.2mm. Acrylic tubes were attached upstream and downstream of the test section, and the former was equipped with a flow straightener array. The test section was placed inside a water-filled acrylic outer tube in order to avoid air-induced magnetic susceptibility gradients. A pipe loop provided flow from pumping hardware outside the MRI scanner room, with the return pipe looping back through an annular gap between the resonator body and the gradient coil inner diameter. Flow was collected in a receiver reservoir, pumped via a variable speed diaphragm pump, fed to a pulsation dampening accumulator, and then back to the test section. Controlled pump bypass enabled very low flow rates whilst keeping the pump oscillation frequency high. Loop circulation timescales are on the order of the scanning timescale. The flow loop was purged of bubbles after filling, and K-Type and alcohol thermometers measurements indicated a fluid (ambient) temperature of 21.8\textsuperscript{$\circ$}C.

% 2. working fluid description
The test solution used here is a 46wt\% haematocrit level blood analogue \citep{Brookshier1993} (0.5wt\% NaCl was omitted because it would interfere with MRI). A 40wt\% glycerine solution in deionised water was first prepared and then used as the solvent for a 0.04wt\% xantham gum solution. The solution appears weakly viscoelastic, with viscous stresses above elastic stresses in their 2Hz oscillatory shear tests, justifying the generalised Newtonian fluid assumption.
% 3. \textbf{}flow-MRI scans (velocity imaging)

Flow-MRI was performed using a Bruker Biospec 70/30USR 7T preclinical scanner (660 mT/m gradients, 35mm ID resonator coil). Images were acquired with uniform radial pixel spacing of 0.1mm and axial slice thickness of 0.08mm. Four scan repetitions were performed in order to reduce noise ($\sim$15 minutes total scanning time).

\subsection{Flow-MRI data pre-processing}
We use phase-contrast MRI and Hadamard encoding to measure all three components of a three-dimensional velocity field using a single set of four, fully-sampled $\bm{k}$-space scans, $\{{s}\}_{j=1}^4$. For each scan we compute its respective complex image, $w_j$, which is given by
\begin{gather}
w_j \coloneqq \rho_j e^{i\varphi_j} = \mathcal{F}^{-1}s_j \quad,
\label{eq:complex_image_full}
\end{gather}
where $\rho_j$ is the nuclear spin density image, $\varphi_j$ is the phase image, and $\mathcal{F}$ is the Fourier transform.
%\footnote{Because the scans $s_j$ fully-sample $\bm{k}$-space, the mapping $s_j \mapsto w_j$ is a bijection, which implies that we can work with the complex images directly, instead of the $k$-space scans (this would not hold true if we had sparsely-sample $k$-space).}. 
The velocity components $u_i$, for $i=1,\dots3$, are then given by
\begin{gather}
u_i = c_i~h_{ij}~\varphi_j \quad,\quad 
h_{ij}= \begin{pmatrix}
-1 & \hphantom{-}1 & \hphantom{-}1 & -1\\
-1 & \hphantom{-}1 & -1 & \hphantom{-}1\\
-1 & -1 & \hphantom{-}1 & \hphantom{-}1
\end{pmatrix}\quad,\quad j=1\ldots4\quad,
\label{eq:vel_from_phase}
\end{gather}
where repeated indices imply summation, and $c_i$ is a known constant that depends on the gyromagnetic
ratio of hydrogen and the gradient pulse properties. In order to remove any phase shift contributions that are not caused by the flow, we conduct an additional no-flow experiment. That is, we acquire a set of four $\bm{k}$-space scans, $\{\bar{s}\}_{j=1}^4$, for the same geometry and field-of-view, but with zero flow (stagnant fluid). We then obtain the no-flow complex images, $\bar{w}_j$, using equation \eqref{eq:complex_image_full}, and compute the corresponding no-flow velocity images using equation \eqref{eq:vel_from_phase}, such that
$\bar{u}_i = \bar{c}_i h_{ij} \bar{\varphi}_j$, where $\bar{c}_i$ is the no-flow constant, which is known. The corrected velocity is then given by $u_i = c_i\theta(h_{ij}\varphi_j) - \bar{c}_i\theta(h_{ij}\bar{\varphi}_j)$, where $\theta(x) \coloneqq x - 2\pi\big(\big\lfloor (\lfloor{x}/\pi \rfloor -1)/2\big\rfloor + 1\big)$ is the phase difference unwrapping function, and $\lfloor{\cdot}/\cdot \rfloor$ denotes integer division. To increase the signal-to-noise ratio (SNR) of steady flow images we acquire $n$ sets (in this study $n=4$) of \mbox{$\bm{k}$-space} scans, generate their respective velocity images $\{u_i\}_{k=1}^{n}$, and compute the average velocity image $\sum_{k=1}^{n} (u_i)_k/n$. The noise variance in the averaged velocity images then reduces to $\sigma^2/n$, where $\sigma^2$ is the noise variance of each individual velocity image. We straighten and centre the averaged flow-MRI images, and, since the flow is axisymmetric, we mirror-average the images to further increase SNR and enforce mirror-symmetry. We generate a mask for the region of interest by segmenting the mirror-averaged nuclear spin density image, and apply this mask to the velocity images. 
Because we solve an inverse N--S problem in a 3D discrete space comprised of trilinear finite elements (voxels), the final pre-processing step is to $L^2$-project the 2D axisymmetric flow-MRI images, $(u_r,u_z)$, to their corresponding 3D flow field, $\bm{u}^\star=(u^\star_x,u^\star_y,u^\star_z)$. Note that the 3D data that we generate \citep{kontogiannis2024_data} have the same (2D) spatial resolution as the 2D images.

\section{Joint flow field reconstruction and Carreau parameter learning}
\label{sec:joint_rec_and_learning}
We apply algorithm \ref{algo_learning_rheo} to the non-Newtonian axisymmetric jet in order to jointly reconstruct the velocity field and learn the rheological parameters of the Carreau fluid. We use the velocimetry data \citep{kontogiannis2024_data}, $\bm{u}^\star$, and compute the data noise covariance, $\datacov = \sigma^2 \mathrm{I}$, where $\mathrm{I}$ is the identity operator, and $\sigma = 0.234$ cm/s \citep{Gudbjartsson1995} 
% \footnote{For fully-sampled $\bm{k}$-space signals and $\text{SNR}>3$, the noise in the flow-MRI images is white and Gaussian \cite{Gudbjartsson1995}.}. 
(for reference, peak jet velocity is $\sim$24 cm/s.) We fix the geometry, $\Omega$, which is known (FDA nozzle), and the outlet b.c. to $\bm{g}_o \equiv \bm{0}$, and infer the unknown Carreau parameters and the inlet b.c., $\bm{g}_i$. To test the robustness of algorithm \ref{algo_learning_rheo}, we assume high uncertainty in the priors by setting the prior mean and covariance of the Carreau parameters to $\bar{\bm{p}}_\mu = \big(\mu_\infty,\delta\mu,\lambda,n\big) = \big(4\cdot10^{-3},10^{-1},5,1\big)$, $\mathcal{C}_{\bar{\bm{p}}_\mu} = \mathrm{diag}\big(0.5\cdot10^{-3},0.5\cdot10^{-1},1,0.5\big)^2$, in SI units, and $\rho = 1099.3$ kg/m\textsuperscript{3}. Note that the prior mean corresponds to a Newtonian fluid with viscosity \mbox{$\mu_e(\bar{\bm{p}}_{\mu}) \equiv \bar{\mu}_\infty+\bar{\delta\mu} \simeq 0.1$ Pa.s}. We set the prior mean of the inlet b.c. to \mbox{$\bar{\bm{g}}_i = (\mathcal{S}^*\bm{u}^\star)|_{\Gamma_i}$}, \emph{i.e.} the restriction of the $\mathcal{S}^*$-projected data on $\Gamma_i$, and the prior covariance to $\mathcal{C}_{\bar{\bm{g}}_i} = \sigma_{\bar{\bm{g}}_i}^2\mathrm{I}$, where \mbox{$\sigma_{\bar{\bm{g}}_i} = 1$ cm/s}. We infer the inlet b.c., instead of fixing its value to $(\mathcal{S}^*\bm{u}^\star)|_{\Gamma_i}$, in order to compensate for measurement noise and local imaging artefacts/biases on (or near) $\Gamma_i$. 
 
\subsection{Flow field reconstruction}
\begin{figure}
\centering
\begin{subfigure}{\figwidth\linewidth}
\centering
\includegraphics[width=.95\linewidth,trim={0cm 0 0 0},clip]{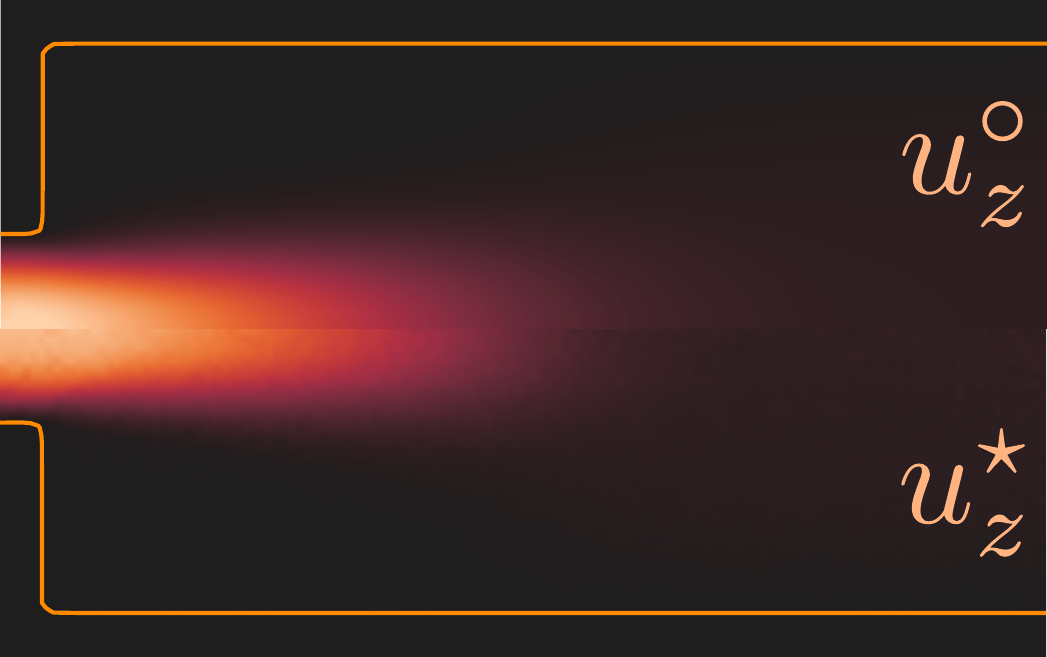}
\caption{axial velocity, $u_z$}
\label{fig:u_z_img}
\end{subfigure}%
\begin{subfigure}{\figwidth\linewidth}
\centering
\includegraphics[width=.95\linewidth,trim={0cm 0 0 0},clip]{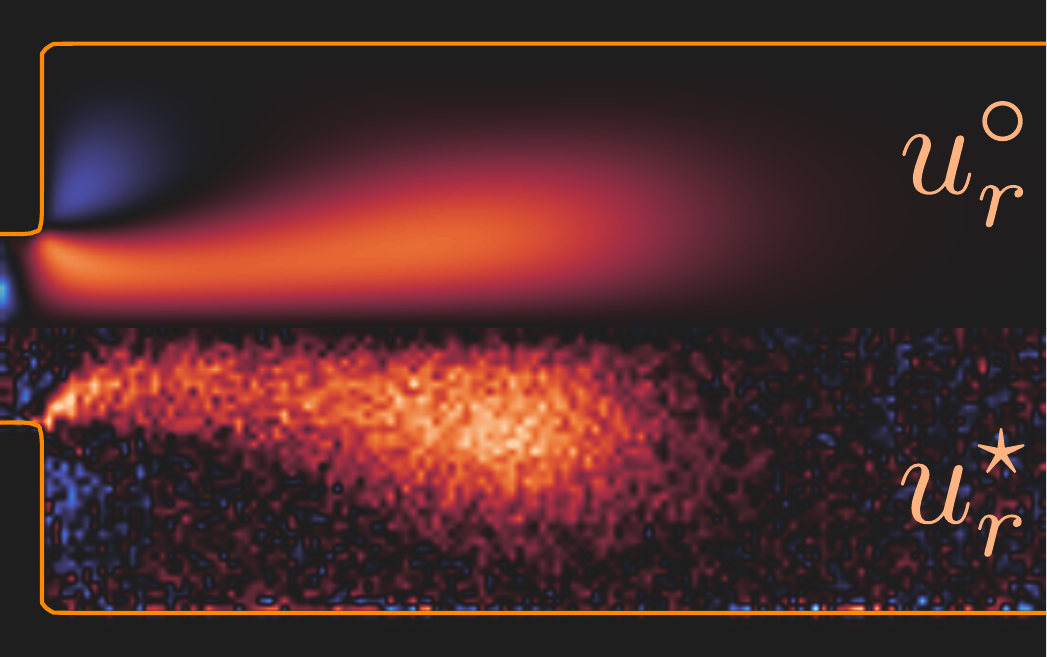}
\caption{radial velocity, $u_r$}
\label{fig:u_r_img}
\end{subfigure}
\begin{subfigure}{\figwidth\linewidth}
\centering
\includegraphics[width=.95\linewidth,trim={0cm 0 0 0},clip]{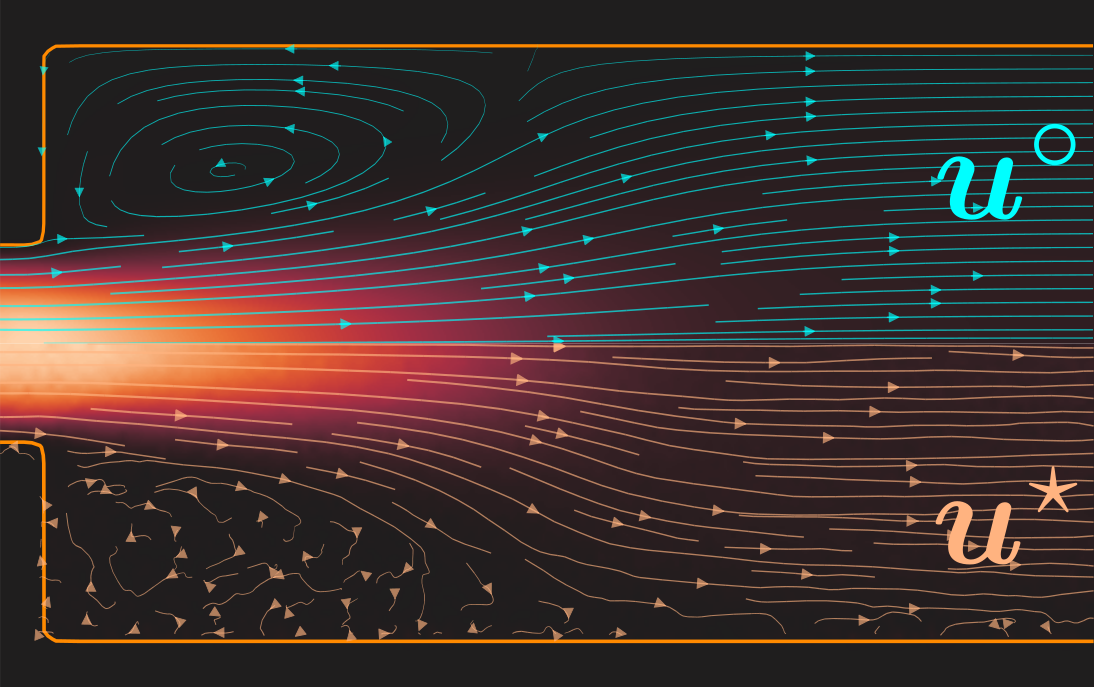}
\caption{streamlines and magnitude}
\label{fig:mag_and_stream}
\end{subfigure}%
\begin{subfigure}{\figwidth\linewidth}
\centering
\includegraphics[width=.95\linewidth,trim={0cm 0 0 0},clip]{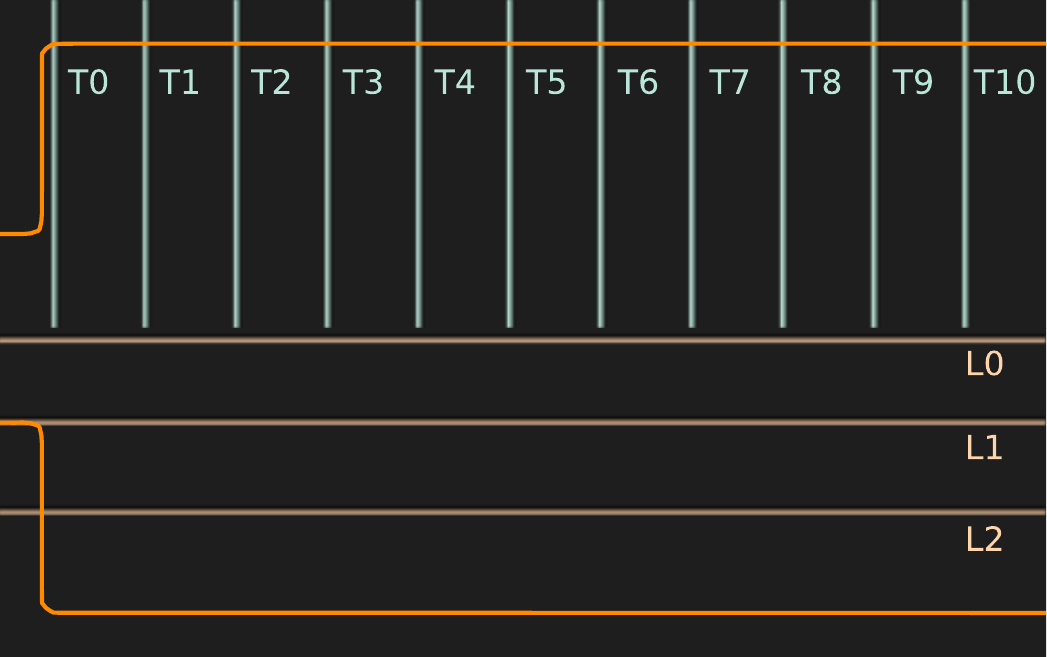}
\caption{slices indices}
\label{fig:slices_map}
\end{subfigure}
\begin{subfigure}{\figwidth\linewidth}
\centering
\includegraphics[width=.95\linewidth,trim={0cm 0 0 0},clip]{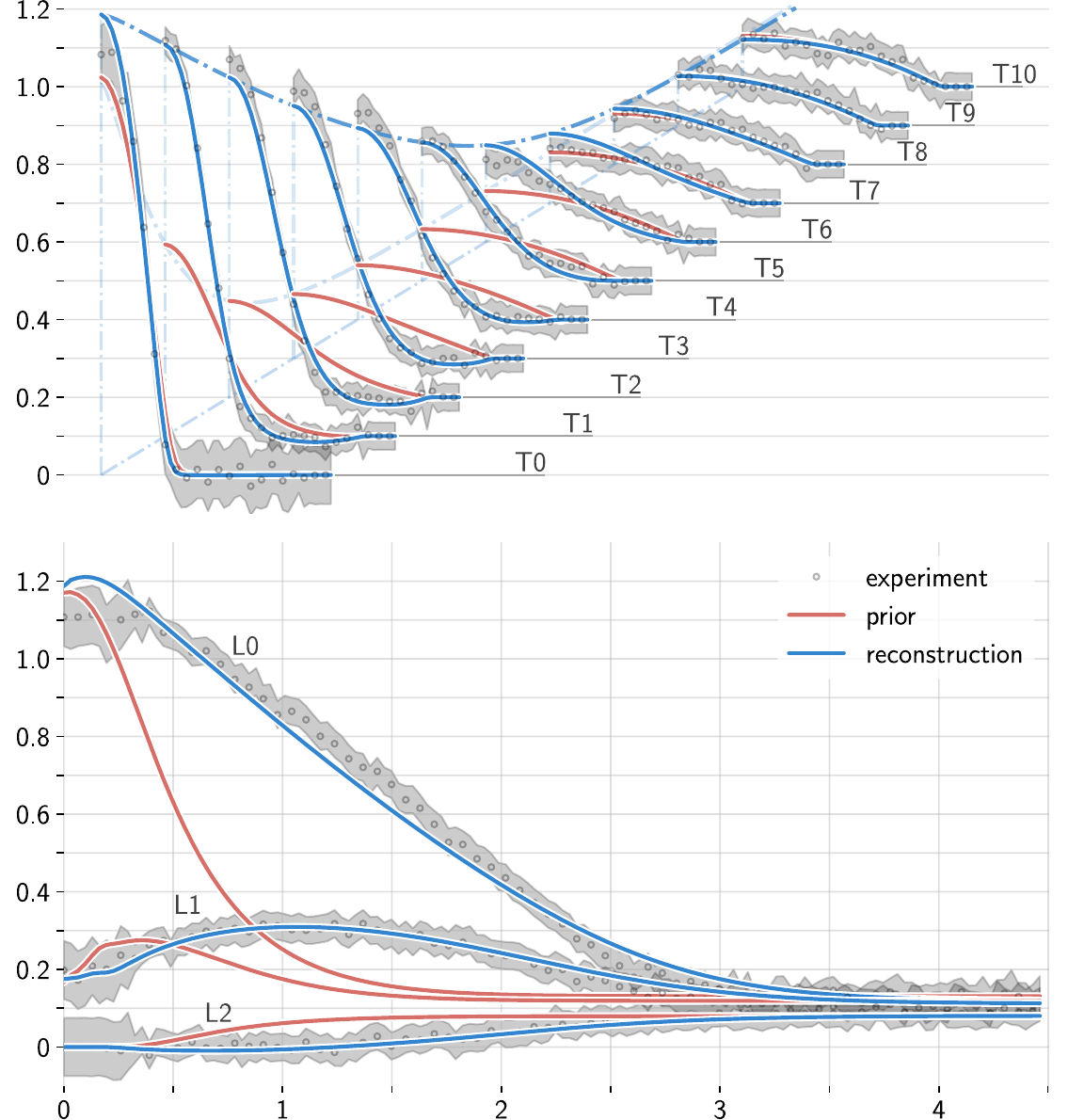}
\caption{axial velocity, $u_z$}
\label{fig:u_z_slices}
\end{subfigure}%
\begin{subfigure}{\figwidth\linewidth}
\centering
\includegraphics[width=.95\linewidth,trim={0cm 0 0 0},clip]{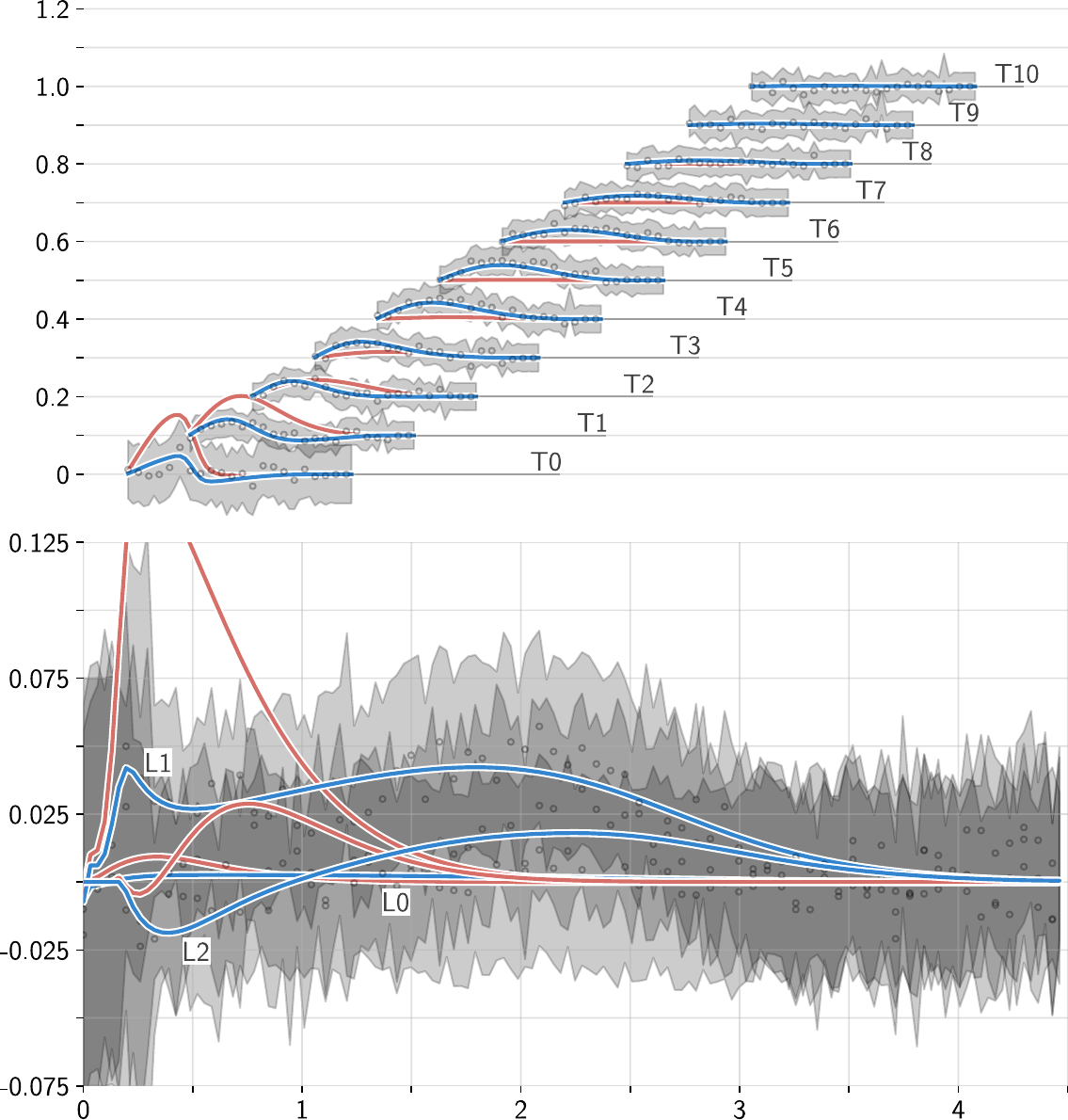}
\caption{radial velocity, $u_r$}
\label{fig:u_r_slices}
\end{subfigure}
\caption{Images and slices of reconstructed (MAP) flow, $\bm{u}^\circ$, vs. velocimetry data, $\bm{u}^\star$. 
% The outline of the geometry is shown in figures \ref{fig:u_z_img}-\ref{fig:slices_map} in orange color. 
In figures \ref{fig:u_z_slices} and \ref{fig:u_r_slices}, velocity is normalised by $U=20$ cm/s, and length is normalised by $L=5$ mm. We separate the transverse slices in the plot by applying a vertical offset of $0.1n$ to the $n$-th slice (the horizontal offset value is immaterial).}
\label{fig:flow_field_reconstruction}
\end{figure}

The reconstructed flow field, $\bm{u}^\circ$, which is generated using algorithm \ref{algo_learning_rheo}, is shown in figure \ref{fig:flow_field_reconstruction} vs. the velocimetry data, $\bm{u}^\star$. We define the average data-model distance by \mbox{$\mathcal{E}(u_\square) \coloneqq \abs{\Omega}^{-1}\norm{u^\star_\square-\mathcal{S}u_\square}_{L^2(\Omega)}$}, where $\abs{\Omega}$ is the volume of $\Omega$, and $\square$ is a symbol placeholder. For the reconstructed velocity field, $\bm{u}^\circ$, we then find $\mathcal{E}(u^\circ_x) = \mathcal{E}(u^\circ_y)= 0.71\sigma$, $\mathcal{E}(u^\circ_z) = 1.40\sigma$, and compare this to the distance between the initial guess, $\bm{u}^{(0)}$, and the data 
\mbox{$\mathcal{E}(u^{(0)}_x) = \mathcal{E}(u^{(0)}_y) = 1.39\sigma$}, $\mathcal{E}(u^{(0)}_z) = 5.87\sigma$.
The inferred (MAP) vs. prior strain-rate magnitude, $\dot{\gamma}$, and effective viscosity field, $\mu_e(\dot{\gamma})$, are shown in figure \ref{fig:gamma_and_mu_e}. Note that we initialise algorithm \ref{algo_learning_rheo} using the prior means, and thus $\mu_e^{(0)} = \mu_e\big(\dot{\gamma}(\bm{u}_0);\bar{\bm{p}}_{\mu}\big) \simeq 0.1$ Pa.s. {\color{black}Using the reconstructed flow field, $\bm{u}^\circ$, and the inferred effective viscosity field, $\mu_e^\circ$, we find that the generalised Reynolds number of this flow is $\Rey_g = 37.5$, where $\Rey_g \coloneqq \rho U_cL_c/\mu_c$, $U_c\coloneqq Q/A = 11.1$ cm/s, $Q$ is the volumetric flow rate, $A$ is the cross-section area before the expansion, $L_c = 4$ mm, and $\mu_c=13$ mPa is the value of the effective viscosity on the wall, before the expansion.}
\subsection{Carreau parameter learning}
\label{sec:carreau_param_learning}
\begin{figure}
\centering
\begin{subfigure}[t]{\figwidth\linewidth}
\centering
\includegraphics[width=.95\linewidth,trim={0cm 0 0 0},clip]{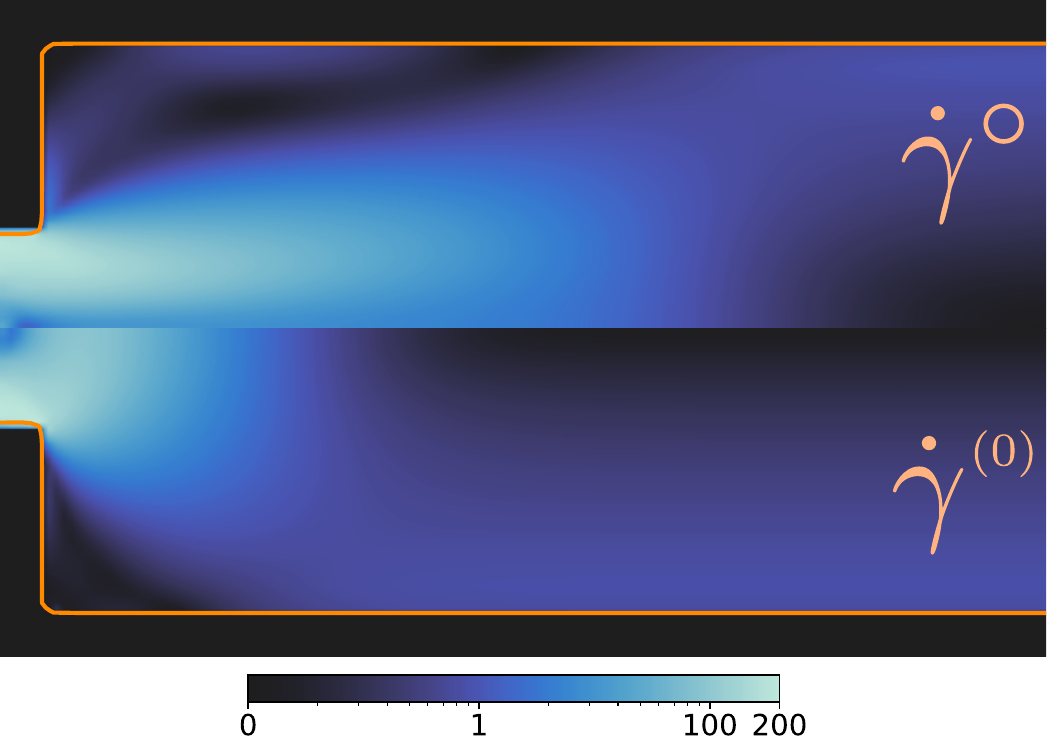}
\caption{strain-rate magnitude [s\textsuperscript{-1}]}
\label{fig:gamma_init_and_rec}
\end{subfigure}
\begin{subfigure}[t]{\figwidth\linewidth}
\centering
\includegraphics[width=.95\linewidth,trim={0cm 0 0 0},clip]{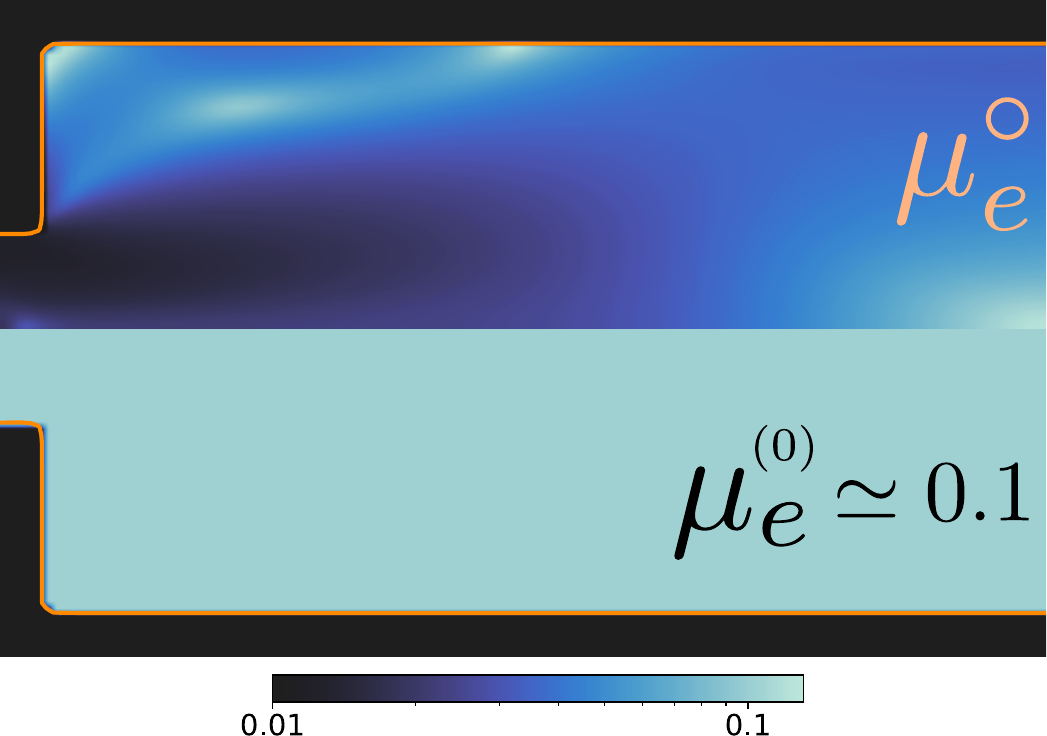}
\caption{effective viscosity [Pa.s]}
\label{fig:mu_e_init_and_rec}
\end{subfigure}
\caption{Inferred (MAP) vs. prior strain-rate magnitude, $\dot{\gamma}$, and effective viscosity, $\mu_e$. }
\label{fig:gamma_and_mu_e}
\end{figure}

\begin{figure}
\centering
\begin{subfigure}[t]{\figwidth\linewidth}
\centering
\includegraphics[width=.95\linewidth]{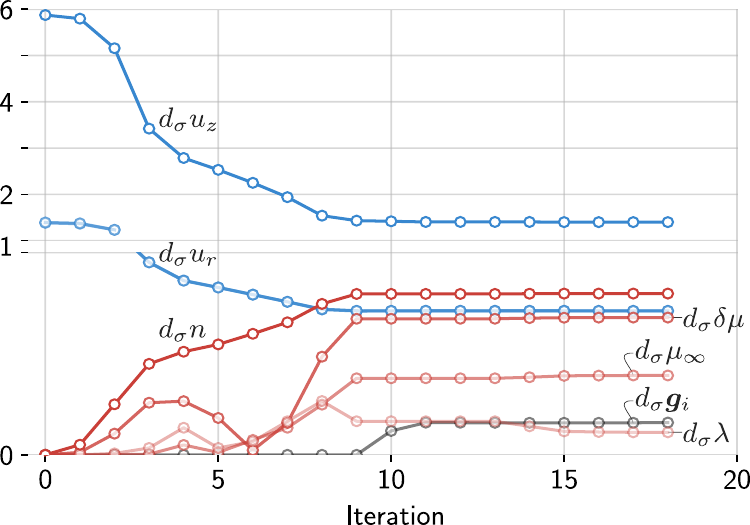}
\caption{optimisation log}
\label{fig:optim_log}
\end{subfigure}%
\begin{subfigure}[t]{\figwidth\linewidth}
\includegraphics[width=.95\linewidth]{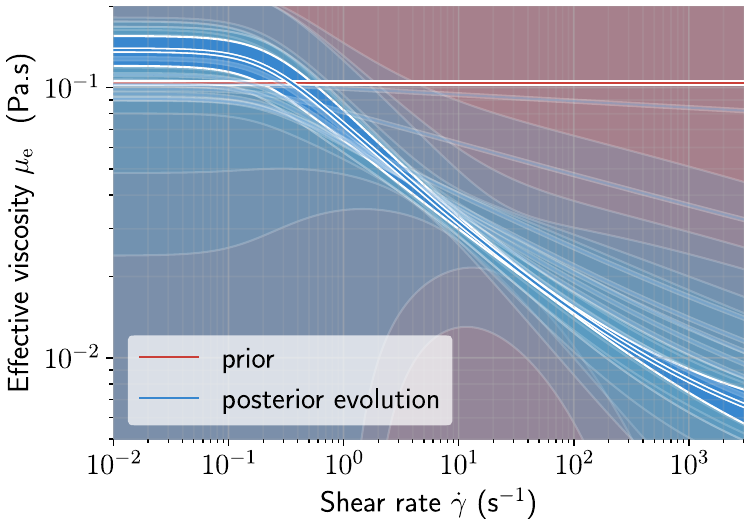}
\caption{p.d.f. evolution of $\mu_e$}
\label{fig:mu_e_history}
\end{subfigure}
\begin{subfigure}[t]{\linewidth}
\centering
\includegraphics[width=\figwidthh\linewidth]{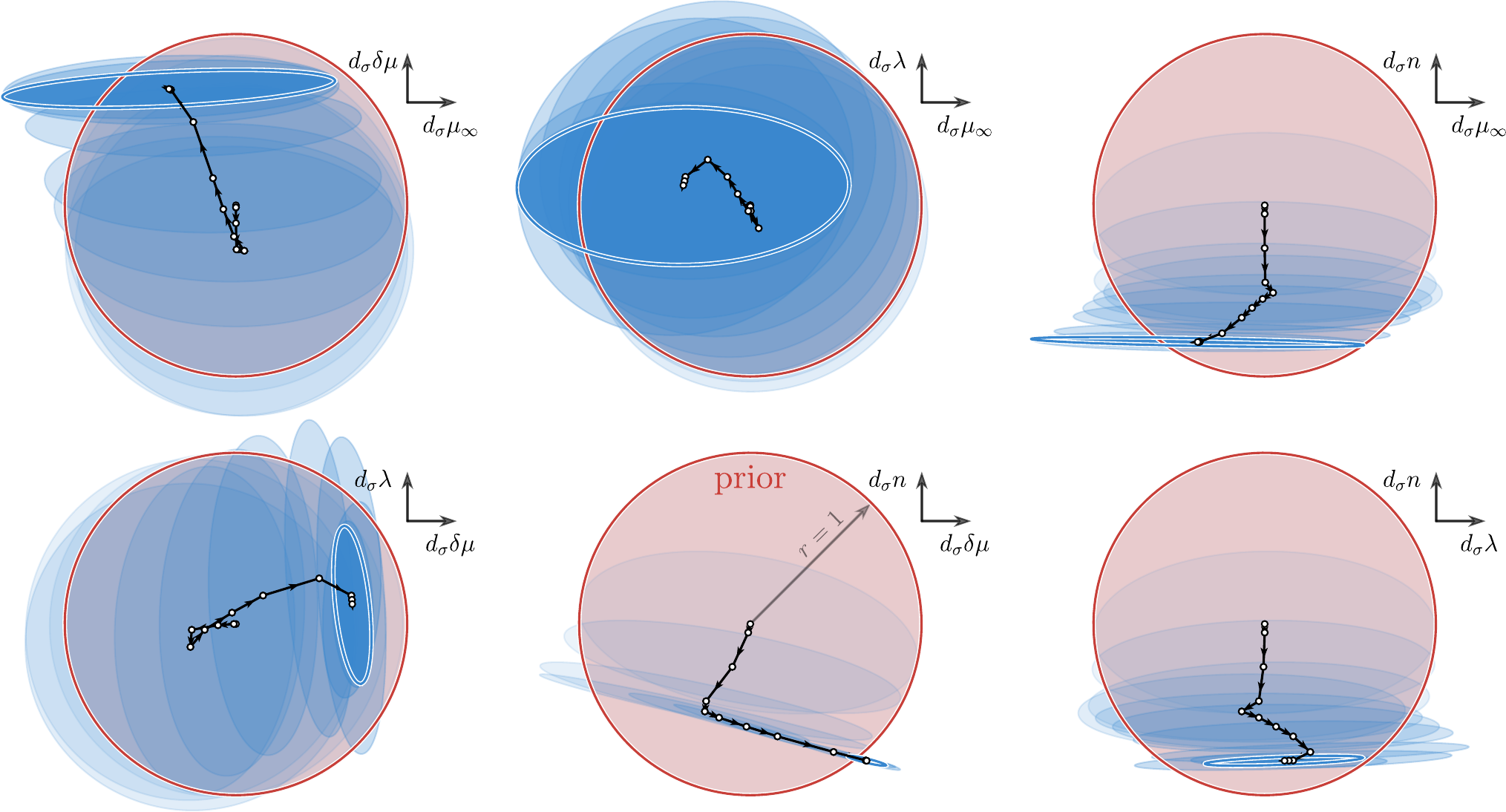}
\caption{p.d.f. evolution of $\bm{p}_\mu$}
\label{fig:mu_e_param_history}
\end{subfigure}
\caption{Optimisation log (figure \ref{fig:optim_log}), and posterior p.d.f. evolution of the effective viscosity (figure \ref{fig:mu_e_history}) and the Carreau parameters (figure \ref{fig:mu_e_param_history}). In figure \ref{fig:mu_e_param_history} the axes are such that $d_\sigma x \coloneqq (x-\bar{x})/\sigma_{\bar{x}}$, where $\bar{x}$ is the prior mean, and $\sigma_{\bar{x}}$ is the prior standard deviation.} 
\end{figure}

According to the optimisation log (figure \ref{fig:optim_log}), the algorithm learns the unknown N--S parameters (\emph{i.e.} the Carreau parameters and the inlet b.c.) in $\sim$20 iterations, but most of the work is done in $\sim$10 iterations. 
Using the Carreau parameters learned at every step, $k$, of the optimisation process, we plot the evolution of the posterior p.d.f. of $\mu_e$ (mean, $\mu_e^{(k)}$, and covariance, $\mathcal{C}^{(k)}_{\mu_e}$) in figure \ref{fig:mu_e_history}. The posterior covariance of $\mu_e^{(k)}$ is given by 
$$\mathcal{C}^{(k)}_{\mu_e} \coloneqq \mathcal{G}^{(k)}_{\mu_e}~\widetilde{\mathcal{C}}^{(k)}_{\bm{p}_{\mu}}~\big(\mathcal{G}^{(k)}_{\mu_e}\big)^*\quad,$$ where $\mathcal{G}^{(k)}_{\mu_e}$ is the Jacobian of the Carreau fluid model \eqref{eq:carreau_model} with respect to its parameters, $\bm{p}_{\mu}$, and $\widetilde{\mathcal{C}}^{(k)}_{\bm{p}_{\mu}}$ is the BFGS approximation of the posterior covariance of the Carreau parameters, ${\mathcal{C}}^{(k)}_{\bm{p}_{\mu}}$. The prior uncertainty, shown in figure \ref{fig:mu_e_history} as a $\pm3\sigma$ red shaded region, is sufficiently high and extends beyond the $\mu_e -\dot{\gamma}$ plotting range\footnote{{\color{black}To ensure that the inversion is stable with respect to prior perturbations, we performed inversions with different priors and found that the inferred posterior parameter distributions were practically the same. A sensitivity analysis with respect to the priors is, however, beyond the scope of the present study.}}. We observe that the posterior uncertainty of $\mu_e$ significantly reduces after assimilating the data in the model, and that the highest uncertainty reduction is for $10 \lessapprox \dot{\gamma} \lessapprox 200$ s\textsuperscript{-1}, which is the $\dot{\gamma}$-range of the laminar jet (see figure \ref{fig:gamma_init_and_rec}). It is worth mentioning that, even though 
we observe a flow for which $\dot{\gamma}\in[0,200]$ s\textsuperscript{-1}, the region that provided the most information is that around the jet because i) inertial effects balance viscous effects, and ii) the local velocity-to-noise ratio is high, hence the uncertainty collapse in the jet-operating $\dot{\gamma}$-range. 

The posterior p.d.f. evolution of the Carreau parameters is shown in figure \ref{fig:mu_e_param_history}. In this case, the prior uncertainty iso-contours can be visualised using hyperellipsoids in $\mathbb{R}^4$ whose centres are $\bar{\bm{p}}_{\mu}$, and axes length proportional to (the columns of) $\mathcal{C}_{\bar{\bm{p}_\mu}}$. To highlight the parameter uncertainty reduction after assimilating the data, we set the origin to $\bar{\bm{p}}_{\mu}$, and scale each dimension of $\mathbb{R}^4$ using (the columns of) $\mathcal{C}_{\bar{\bm{p}}_\mu}$. In this transformed space the prior uncertainty iso-contours are hyperspheres, and the posterior uncertainty iso-contours are hyperellipsoids, whose slices are shown in figure \ref{fig:mu_e_param_history}. It is interesting to note that, after assimilating the velocimetry data, the posterior uncertainty collapses along the axes $n$, $\delta\mu$, whilst it slightly decreases along the axes $\mu_\infty$, $\lambda$. This indicates that there is insufficient information in the data to further collapse the prior uncertainties of $\lambda$ and $\mu_\infty$. 

It is known that for models with univariate design (e.g. shear rate, $\dot{\gamma}$) and observable variables (e.g. effective viscosity, $\mu_e$) in order to learn the most about the unknown parameters (e.g. $\bm{p}_{\mu}$) we need to perform experiments with the design parameters at which the model is most uncertain \citep[Section~4.1]{Yoko_Juniper_2024}. Here, figure \ref{fig:mu_e_history} shows that the model is most uncertain for $\dot{\gamma} \ll 10$ s\textsuperscript{-1} and $\dot{\gamma} \gg 200$ s\textsuperscript{-1}. Consequently, in order to further collapse the uncertainty of $\lambda$ and $\mu_\infty$, we require more (or higher SNR) flow-MRI data for \mbox{$\dot{\gamma} \ll 10$ s\textsuperscript{-1}} and \mbox{$\dot{\gamma} \gg 200$ s\textsuperscript{-1}}. In particular, since we have used flow-MRI data of a jet for which \mbox{$\dot{\gamma}\in[0,200]$ s\textsuperscript{-1}}, we would require new experiments with i) higher SNR, since information at low velocity magnitudes, i.e. \mbox{$\dot{\gamma} \ll 10$ s\textsuperscript{-1}}, is corrupted by noise, and ii) higher velocity magnitudes (for the same geometry), since information at \mbox{$\dot{\gamma} \gg 200$ s\textsuperscript{-1}} is missing from the current experiment.

\subsection{Validation via an independent rheometry experiment}
\begin{figure}
\centering
\begin{subfigure}[b]{\figwidth\linewidth}
\centering
\includegraphics[width=0.95\linewidth,trim={0cm 0 0 0},clip]{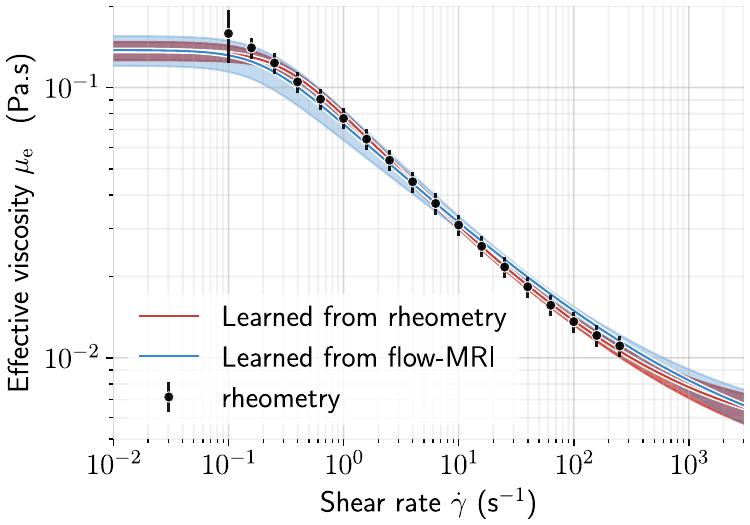}
\caption{learned effective viscosity}
\label{fig:eff_visc}
\end{subfigure}\hspace{0.5cm}
\begin{subfigure}[b]{\figwidth\linewidth}\centering
\begin{adjustbox}{max width=.95\textwidth}
  \begin{tabular}{cccc}
    \multicolumn{4}{c}{\color{gray}priors}\\\arrayrulecolor{gray}\hline\hline\\[-1.5ex]
    \color{gray}$\mu_{\infty}$ (mPa.s) & \color{gray}$\delta\mu$ (mPa.s) & \color{gray}$\lambda$ & \color{gray}$n$ \\[1pt]
  \color{gray}$4.00\pm1.50$ & \color{gray}$100\pm150$ & \color{gray}$5.00\pm3.00$ & \color{gray}$1.00\pm1.50$\\[2ex]
  \multicolumn{4}{c}{learned from flow-MRI}\\\hline\hline\\[-1.5ex]
  \color{Blue}$\mu_{\infty}$ (mPa.s) & \color{Blue}$\delta\mu$ (mPa.s) & $\color{Blue}\lambda$ & $\color{Blue}n$ \\[1pt]
  $3.80\pm1.46$ & $134\pm16.8$ & $5.11\pm1.38$ & $0.601\pm0.0437$\\[2ex]
   \multicolumn{4}{c}{learned from rheometry}\\\hline\hline\\[-1.5ex]
  \color{BrickRed}$\mu_{\infty}$ (mPa.s) & \color{BrickRed}$\delta\mu$ (mPa.s) & $\color{BrickRed}\lambda$ & $\color{BrickRed}n$ \\[1pt]
  $4.64\pm1.18$ & $132\pm11.0$ & $3.36\pm0.904$ & $0.539\pm0.0288$\\
  & & &
  \end{tabular}
  \end{adjustbox}
  % \vspace{0.5cm}
  \caption{learned model parameters}
  \label{tab:inferred_params}
\end{subfigure}
\caption{Learned Carreau fit to rheometry data, learned model parameters (MAP estimates), and assumed priors. Uncertainties in the figures correspond to $3\sigma$ intervals.}
\label{fig:visc_model_fitting}
\end{figure}

Steady-shear rheometry of the test solution was conducted using a Netzsch Kinexus rheometer with a $\diameter$27.5mm cup and a $\diameter$25mm bob geometry, at the same temperature as the flow-MRI experiment. The experiment was conducted to validate the Carreau parameters learned from the flow-MRI data (section \ref{sec:carreau_param_learning}) against rheometry data. To find the most likely Carreau parameters that fit the rheometry data, we use Bayesian inversion (see section \ref{sec:learning_rheo}). In this case, operator $\mathcal{Z}$ corresponds to the Carreau model, given by the explicit relation \eqref{eq:carreau_model}, and operator $\mathcal{G}$ corresponds to the Jacobian of the Carreau model with respect to its parameters. We use the same prior as in section \ref{sec:joint_rec_and_learning}. Because the prior uncertainty is sufficiently high relative to the noise variance, the bias it introduces to the model fit is negligible. The Carreau parameters learned from flow-MRI vs. rheometry are shown in figure \ref{fig:visc_model_fitting}. We observe that the parameters learned from flow-MRI agree well with rheometry data, considering uncertainties (table \ref{tab:inferred_params}), and that the learned effective viscosity field fits the rheometry data (figure \ref{fig:eff_visc}). As in the case of learning from flow-MRI, it is not possible to infer $\lambda$ and $\mu_\infty$ with high certainty when data, $\mu_e(\dot{\gamma})$, for $\dot{\gamma} \ll 10$ s\textsuperscript{-1} and $\dot{\gamma} \gg 200$ s\textsuperscript{-1} are missing (or the measurement uncertainty is high).

\section{Summary and conclusions}
We have formulated a Bayesian inverse N--S problem that assimilates velocimetry data of 3D steady incompressible flow in order to jointly reconstruct the flow field and learn the unknown N--S parameters. By incorporating a Carreau shear-thinning viscosity model into the N--S problem, we devise an algorithm that learns the Carreau parameters of a shear-thinning fluid, and estimates their uncertainties, from velocimetry data alone. Then we conduct a flow-MRI experiment to obtain velocimetry data of an axisymmetric laminar jet through an idealised medical device (FDA nozzle), for a blood analogue fluid. We show that the algorithm successfully reconstructs the noisy flow field, and, at the same time, learns the Carreau parameters and their uncertainties. To ensure that the learned Carreau parameters explain the rheology of the fluid, instead of simply fitting the velocimetry data, we conduct an additional rheometry experiment. We find that the Carreau parameters learned from the flow-MRI data alone are in very good agreement with the parameters learned from the rheometry experiment (taking into account their uncertainties), and that the learned effective viscosity field fits the rheometry data. In this paper we have applied the algorithm to a Carreau fluid. The present algorithm, however, accepts any generalised Newtonian fluid, as long as the model is (weakly) differentiable. More complicated non-Newtonian behaviour, such as viscoelasticity, can be learned from velocimetry data if a viscoelastic model (e.g., Oldroyd-B fluid) is incorporated into the N--S problem. {\color{black} The present study has demonstrated the method on a single dataset. In future work we will extend this to more test cases and different experimental configurations.}

\backsection[Acknowledgement]{Thanks to \emph{Matthew Yoko} (Cambridge) for useful discussions on optimal experiment design, and for providing the script that straightens and centres axisymmetric object images.}

\backsection[Funding]{The authors are supported by EPSRC National Fellowships in Fluid Dynamics (NFFDy), grants EP/X028232/1 (AK), EP/X028089/1 (RH), and EP/X028321/1 (ELM).}

\backsection[Declaration of interests]{The authors report no conflict of interest.}

\backsection[Data availability statement]{The experimental flow-MRI data used in this study are available at \url{https://doi.org/10.17863/CAM.113320}.}

% \backsection[Author contributions]{\textbf{Alexandros Kontogiannis:} conceptualization, formal analysis, funding acquisition, investigation, methodology, project administration, software, validation, visualization, writing -- original draft, writing -- review \& editing. \textbf{Richard Hodgkinson:} conceptualization, data curation, investigation, funding acquisition, project administration, resources, writing -- original draft, writing -- review \& editing. \textbf{Steven Reynolds:} data curation, investigation. \textbf{Emily L. Manchester:} conceptualization, funding acquisition, investigation, project administration, software, validation, writing -- original draft, writing -- review \& editing}

\bibliographystyle{jfm}
\bibliography{references}

\begin{thebibliography}{24}
\expandafter\ifx\csname natexlab\endcsname\relax\def\natexlab#1{#1}\fi
\def\au#1{#1} \def\ed#1{#1} \def\yr#1{#1}\def\at#1{#1}\def\jt#1{\textit{#1}} \def\bt#1{#1}\def\bvol#1{\textbf{#1}} \def\vol#1{#1} \def\pg#1{#1} \def\publ#1{#1}\def\arxiv#1{#1}\def\org#1{#1}\def\st#1{\textit{#1}}

\bibitem[Brookshier \& Tarbell(1993)]{Brookshier1993}
{\sc \au{Brookshier, K.~A.} \& \au{Tarbell, J.~M.}} \yr{1993}  \at{{Evaluation of a transparent blood analog fluid: Aqueous Xanthan gum/glycerin}}.  \jt{Biorheology}  \bvol{30},  \pg{107--116}, 2.

\bibitem[Elkins \& Alley(2007)]{Elkins2007}
{\sc \au{Elkins, C.~J.} \& \au{Alley, M.~T.}} \yr{2007}  \at{{Magnetic resonance velocimetry: applications of magnetic resonance imaging in the measurement of fluid motion}}.  \jt{Experiments in Fluids}  \bvol{43}~(6),  \pg{823--858}.

\bibitem[Freund \& Ewoldt(2015)]{10.1122/1.4915299}
{\sc \au{Freund, Jonathan~B.} \& \au{Ewoldt, Randy~H.}} \yr{2015}  \at{{Quantitative rheological model selection: Good fits versus credible models using Bayesian inference}}.  \jt{Journal of Rheology}  \bvol{59}~(3),  \pg{667--701},  \arxiv{arXiv: https://pubs.aip.org/sor/jor/article-pdf/59/3/667/16014411/667\_1\_online.pdf}.

\bibitem[Goldfarb {\em et~al.\/}(2020)Goldfarb, Ren \& Bahamou]{Goldfarb2020}
{\sc \au{Goldfarb, D.}, \au{Ren, Y.} \& \au{Bahamou, A.}} \yr{2020} {Practical Quasi-Newton Methods for Training Deep Neural Networks}.  \bt{In {\em Advances in Neural Information Processing Systems\/} (ed. \ed{H.~Larochelle, M.~Ranzato, R.~Hadsell, M.F. Balcan \& H.~Lin})}, ,  \vol{vol.~33},  \pg{pp. 2386--2396}.  \publ{Curran Associates, Inc.}

\bibitem[Gudbjartsson \& Patz(1995)]{Gudbjartsson1995}
{\sc \au{Gudbjartsson, H.} \& \au{Patz, S.}} \yr{1995}  \at{{The Rician Distribution of Noisy MRI Data}}.  \jt{Magnetic Resonance in Medicine}  \bvol{34}~(6),  \pg{910--914}.

\bibitem[Hariharan {\em et~al.\/}(2011)Hariharan, Giarra, Reddy, Day, Manning, Deutsch, Stewart, Myers, Berman, Burgreen, Paterson \& Malinauskas]{fda_nozzle}
{\sc \au{Hariharan, P.}, \au{Giarra, M.}, \au{Reddy, V.}, \au{Day, S.~W.}, \au{Manning, K.~B.}, \au{Deutsch, S.}, \au{Stewart, S. F.~C.}, \au{Myers, M.~R.}, \au{Berman, M.~R.}, \au{Burgreen, G.~W.}, \au{Paterson, E.~G.} \& \au{Malinauskas, R.~A.}} \yr{2011}  \at{{Multilaboratory Particle Image Velocimetry Analysis of the FDA Benchmark Nozzle Model to Support Validation of Computational Fluid Dynamics Simulations}}.  \jt{Journal of Biomechanical Engineering}  \bvol{133}~(4),  \pg{041002}.

\bibitem[Konigsberg {\em et~al.\/}(2013)Konigsberg, Nicholson, Halley, Kealy \& Bhattacharjee]{KonigsbergNicholsonHalleyKealyBhattacharjee+2013}
{\sc \au{Konigsberg, D.}, \au{Nicholson, T.~M.}, \au{Halley, P.~J.}, \au{Kealy, T.~J.} \& \au{Bhattacharjee, P.~K.}} \yr{2013}  \at{Online process rheometry using oscillatory squeeze flow}.  \jt{Applied Rheology}  \bvol{23}~(3),  \pg{35688}.

\bibitem[Kontogiannis {\em et~al.\/}(2024{\natexlab{{\em a\/}}})Kontogiannis, Elgersma, Sederman \& Juniper]{Kontogiannis2024}
{\sc \au{Kontogiannis, Alexandros}, \au{Elgersma, Scott~V}, \au{Sederman, Andrew~J} \& \au{Juniper, Matthew}} \yr{2024{\natexlab{{\em a\/}}}}  \at{{Bayesian inverse Navier-Stokes problems: joint flow field reconstruction and parameter learning}}.  \jt{Inverse Problems} .

\bibitem[Kontogiannis {\em et~al.\/}(2022)Kontogiannis, Elgersma, Sederman \& Juniper]{Kontogiannis2021}
{\sc \au{Kontogiannis, A.}, \au{Elgersma, S.~V.}, \au{Sederman, A.~J.} \& \au{Juniper, M.~P.}} \yr{2022}  \at{{Joint reconstruction and segmentation of noisy velocity images as an inverse Navier--Stokes problem}}.  \jt{Journal of Fluid Mechanics}  \bvol{944},  \pg{A40}.

\bibitem[Kontogiannis {\em et~al.\/}(2024{\natexlab{{\em b\/}}})Kontogiannis, Hodgkinson, Reynolds \& Manchester]{kontogiannis2024_data}
{\sc \au{Kontogiannis, Alexandros}, \au{Hodgkinson, Richard}, \au{Reynolds, Steven} \& \au{Manchester, Emily}} \yr{2024{\natexlab{{\em b\/}}}}  \at{{Research data supporting: `Learning rheological parameters of non-Newtonian fluids from velocimetry data'}} .

\bibitem[Kontogiannis \& Juniper(2023)]{Kontogiannis2022b}
{\sc \au{Kontogiannis, A.} \& \au{Juniper, M.~P.}} \yr{2023}  \at{{Physics-Informed Compressed Sensing for PC-MRI: An Inverse Navier-Stokes Problem}}.  \jt{IEEE Transactions on Image Processing}  \bvol{32},  \pg{281--294}.

\bibitem[MacKay(2003)]{MacKay2003}
{\sc \au{MacKay, David J.~C.}} \yr{2003} {\em {Information Theory, Inference and Learning Algorithms}\/}.  \publ{Cambridge University Press}.

\bibitem[Noto {\em et~al.\/}(2023)Noto, Ohie, Yoshida \& Tasaka]{Noto2023}
{\sc \au{Noto, Daisuke}, \au{Ohie, Kohei}, \au{Yoshida, Taiki} \& \au{Tasaka, Yuji}} \yr{2023}  \at{Optical spinning rheometry test on viscosity curves of less viscous fluids at low shear rate range}.  \jt{Experiments in Fluids}  \bvol{64}~(1),  \pg{18}.

\bibitem[Ohie {\em et~al.\/}(2022)Ohie, Yoshida, Tasaka, Sugihara-Seki \& Murai]{Ohie2022}
{\sc \au{Ohie, Kohei}, \au{Yoshida, Taiki}, \au{Tasaka, Yuji}, \au{Sugihara-Seki, Masako} \& \au{Murai, Yuichi}} \yr{2022}  \at{Rheological characterization and flow reconstruction of polyvinylpyrrolidone aqueous solutions by means of velocity profiling-based rheometry}.  \jt{Experiments in Fluids}  \bvol{63}.

\bibitem[Ranftl {\em et~al.\/}(2023)Ranftl, Müller, Windberger, Brenn \& von~der Linden]{Ranftl2023}
{\sc \au{Ranftl, S.}, \au{Müller, T.~S.}, \au{Windberger, U.}, \au{Brenn, G.} \& \au{von~der Linden, W.}} \yr{2023}  \at{{A Bayesian approach to blood rheological uncertainties in aortic hemodynamics}}.  \jt{International Journal for Numerical Methods in Biomedical Engineering}  \bvol{39}.

\bibitem[Sequeira \& Janela(2007)]{Sequeira2007}
{\sc \au{Sequeira, A.} \& \au{Janela, J.}} \yr{2007} {\em {An overview of some mathematical models of blood rheology}\/}.  \publ{Springer Netherlands}.

\bibitem[Shwetank {\em et~al.\/}(2022)Shwetank, Gerhard, Sunil, Asad \& Krishna]{Krishna2022}
{\sc \au{Shwetank, K.}, \au{Gerhard, T.}, \au{Sunil, K.}, \au{Asad, E.} \& \au{Krishna, R.}} \yr{2022}  \at{Ultrasound velocity profiling technique for in-line rheological measurements: A prospective review}.  \jt{Measurement}  \bvol{205},  \pg{112152}.

\bibitem[Stewart {\em et~al.\/}(2012)Stewart, Paterson, Burgreen, Hariharan, Giarra, Reddy, Day, Manning, Deutsch, Berman, Myers \& Malinauskas]{Stewart2012}
{\sc \au{Stewart, Sandy F.~C.}, \au{Paterson, Eric~G.}, \au{Burgreen, Greg~W.}, \au{Hariharan, Prasanna}, \au{Giarra, Matthew}, \au{Reddy, Varun}, \au{Day, Steven~W.}, \au{Manning, Keefe~B.}, \au{Deutsch, Steven}, \au{Berman, Michael~R.}, \au{Myers, Matthew~R.} \& \au{Malinauskas, Richard~A.}} \yr{2012}  \at{{Assessment of CFD Performance in Simulations of an Idealized Medical Device: Results of FDA's First Computational Interlaboratory Study}}.  \jt{Cardiovascular Engineering and Technology}  \bvol{3}~(2),  \pg{139--160}.

\bibitem[Tarantola(2005)]{Tarantola2005}
{\sc \au{Tarantola, A.}} \yr{2005} {\em {Inverse Problem Theory and Methods for Model Parameter Estimation}\/}.  \publ{SIAM}.

\bibitem[Tasaka {\em et~al.\/}(2021)Tasaka, Yoshida \& Murai]{Tasaka2021}
{\sc \au{Tasaka, Yuji}, \au{Yoshida, Taiki} \& \au{Murai, Yuichi}} \yr{2021}  \at{Nonintrusive in-line rheometry using ultrasonic velocity profiling}.  \jt{Industrial \& Engineering Chemistry Research}  \bvol{60}~(30),  \pg{11535--11543},  \arxiv{arXiv: https://doi.org/10.1021/acs.iecr.1c01795}.

\bibitem[Worthen {\em et~al.\/}(2014)Worthen, Stadler, Petra, Gurnis \& Ghattas]{Worthen2014}
{\sc \au{Worthen, J.}, \au{Stadler, G.}, \au{Petra, N.}, \au{Gurnis, M.} \& \au{Ghattas, O.}} \yr{2014}  \at{{Towards adjoint-based inversion for rheological parameters in nonlinear viscous mantle flow}}.  \jt{Physics of the Earth and Planetary Interiors}  \bvol{234},  \pg{23--34}.

\bibitem[Yoko \& Juniper(2024{\natexlab{{\em a\/}}})]{Yoko_Juniper_2024b}
{\sc \au{Yoko, Matthew} \& \au{Juniper, Matthew~P.}} \yr{2024{\natexlab{{\em a\/}}}}  \at{Adjoint-accelerated bayesian inference applied to the thermoacoustic behaviour of a ducted conical flame}.  \jt{Journal of Fluid Mechanics}  \bvol{985},  \pg{A38}.

\bibitem[Yoko \& Juniper(2024{\natexlab{{\em b\/}}})]{Yoko_Juniper_2024}
{\sc \au{Yoko, Matthew} \& \au{Juniper, Matthew~P.}} \yr{2024{\natexlab{{\em b\/}}}}  \at{{Optimal experiment design with adjoint-accelerated Bayesian inference}}.  \jt{Data-Centric Engineering}  \bvol{5},  \pg{e17}.

\bibitem[Yoshida {\em et~al.\/}(2022)Yoshida, Ohie \& Tasaka]{Yoshida2022}
{\sc \au{Yoshida, Taiki}, \au{Ohie, Kohei} \& \au{Tasaka, Yuji}} \yr{2022}  \at{In situ measurement of instantaneous viscosity curve of fluids in a reserve tank}.  \jt{Industrial \& Engineering Chemistry Research}  \bvol{61}~(31),  \pg{11579--11588},  \arxiv{arXiv: https://doi.org/10.1021/acs.iecr.2c01792}.

\end{thebibliography}
\end{document}